\documentclass[twocolumn]{aastex6}
\usepackage{graphicx}
\usepackage{amsmath,amssymb}
\usepackage{units}
\usepackage{color}

\newcommand{\BF}{\begin{figure}\begin{center}}
\newcommand{\EF}{\end{center}\end{figure}}
\newcommand{\BE}{\begin{equation}}
\newcommand{\EE}{\end{equation}}
\newcommand{\BEA}{\begin{eqnarray}}
\newcommand{\EEA}{\end{eqnarray}}
\newcommand{\ti}{\textit}
\newcommand{\tr}{\textrm}

\shorttitle{ALMA Imaging of Jet-ISM Interaction 
in MG\,J0414+0534}
\shortauthors{Inoue, Matsushita, Nakanishi and Minezaki}
\begin{document}


\title{ALMA 50-parsec-resolution Imaging of Jet-ISM Interaction 
in the Lensed Quasar MG\,J0414+0534}

\author{Kaiki Taro Inoue}
\affiliation{Faculty of Science and Engineering, 
Kindai University, Higashi-Osaka, 577-8502, Japan}

\author{Satoki Matsushita}
\affiliation{Institute of Astronomy and Astrophysics, Academia Sinica, \\
11F of Astronomy-Mathematics Building, AS/NTU, No.1, Sec.4, Roosevelt Rd., Taipei 10617, Taiwan, R.O.C.}

\author{Kouichiro Nakanishi}
\affiliation{National Astronomical Observatory of Japan, Mitaka, Tokyo 181-8588, Japan, \\
The Graduate University for Advanced Studies, SOKENDAI, Mitaka, Tokyo 181-8588, Japan}

\author{Takeo Minezaki}
\affiliation{Institute of Astronomy, School of Science, University of
Tokyo, Mitaka, Tokyo 181-0015, Japan}

\begin{abstract}
We report our high-resolution $(0\farcs 03- 0\farcs 07)$ Atacama Large Millimeter/submillimeter Array (ALMA) imaging of
the quadruply lensed radio-loud quasar MG\,J0414+0534 at redshift $z=2.639$
in the continuum and the broad CO(11-10) line at $\sim 340\,$GHz.
With the help of strong lensing magnification and ALMA's high resolution,
we succeeded in resolving the jet/dust and CO gas in the quasar host galaxy
both extending up to $\sim 1\,$kpc, with a resolution of $\sim 50\,$pc for the first time.
Both the continuum emission and the CO(11-10) line have a similar bimodal structure aligned
with the quasar jets ($\sim 200\,$pc) observed by Very Long Baseline Interferometry (VLBI) at
$5\,$GHz and $8.4\,$GHz. The CO gas in the vicinity of both the eastern and western jet components at the location of $\sim 80\,$pc from the quasar core are moving at high velocities, up to
$\pm 600\,$$\textrm{km}\,\textrm{s}^{-1}$ relative to the core. The observed features show clear evidence of strong interaction between the jets and interstellar medium (ISM). High temperature and high-density environments in the ISM of the quasar host galaxy, as suggested from CO line ratios, also support this result. The small scale of the jets, the jet-ISM interaction, and the continuum spectral energy distribution of this source indicate that we are watching the infancy stage of quasar radio activity.

\end{abstract}

\section{Introduction} \label{sec:intro}
Recently, numerous observational studies in interactions between radio jets from active galactic
nuclei (AGNs) and interstellar medium (ISM) have been reported.  However, spatially resolved observations that can clearly see the jet-ISM interaction are still limited to a few objects. One of the nearest (7.1\,Mpc) Seyfert 2 galaxies, M51, 
shows both diffuse (CO) and dense (HCN) molecular gas along the radio jet, with those velocity gradients similar to that of an optical emission line on the jets. In addition, there is a density gradient with denser molecular gas closer to the jet. These observational results indicate the presence of the jet-entrained molecular gas \citep{matsushita2004,matsushita2007,matsushita2015}.
Relatively nearby (47.9~Mpc) radio-loud Seyfert 2 galaxy IC\,5063 shows dense and warm molecular gas components with broad velocity width at the knots of radio jets, suggesting a jet-ISM interaction \citep{tadhunter2014,morganti2015,osterloo2017}. 
The above two examples clearly show the presence of interaction between radio jets and molecular gas. However, such interactions in early universe (i.e., high-z galaxies) have not been studied observationally. Such observations are important to study the evolution of AGNs and ISM.

Gravitationally lensed radio-loud quasars provide us excellent opportunity
for investigating interactions between jet and ambient ISM in high-z galaxies.
Among them, MG\,J0414+0534 \citep{hewitt1992}, a quadruply lensed
radio-loud quasar at redshift $z_\ti{S}=2.639$ \citep{lawrence1995} is one
the most promising targets due to its large magnification and spatially resolved small-scale ($\lesssim 1\,$kpc) radio jet components, observed at 5\,GHz \citep{trotter2000} and 8.4\,GHz \citep{ros2000}.

In our Cycle 2 data \citep{inoue2017}, we serendipitously discovered
a broad CO(11-10) emission line in MG\,J0414+0534 \citep[see also][]{stacey2018}. Such a highly-excited CO line suggests the presence of molecular outflows.
In ALMA Cycle 4, we proposed carrying out long baseline observations of
the CO(11-10) line to probe the distribution and kinematics of molecular
gas surrounding quasar jets on small scales.

In this letter, we present the results of our ALMA Cycle 4 high angular
resolution ($0\farcs03-0\farcs05$) observations (Project ID: 2016.1.00281.S,
PI: S. Matsushita) combined with our Cycle 2 low angular resolution
($0\farcs1-0\farcs3$) observations of the continuum emission and CO(11-10)
line of MG\,J0414+0534 at 0.88\,mm.
In what follows, we adopt a \ti{Planck} 2018
cosmology with matter density $\Omega_{m,0}=0.315$, energy density of 
cosmological constant $\Omega_{\Lambda,0}=0.685$, and Hubble constant
$H_0=67.4$\,$\textrm{km}\,\textrm{s}^{-1}$.

\section{Observation and Reduction}
Our ALMA Cycle 4 observations were carried out on 2017 November 1, 8, 10, and 11.
The numbers of antennas used in the observations were 49, 48, 45, and 49,
and the total on-source integrating times were 38.72, 39.00, 40.65, and 38.88\,minutes, respectively. The maximum and minimum baselines were 13.8944\,km and 113\,m, respectively. The phase center was $\alpha=$04$^\tr{h}$14$^\tr{m}$37$^\tr{s}$.7686, $\delta=$+05$^\circ$34$'$42\farcs 352 (J2000). The ALMA correlator was configured to have four spectral windows centered at 335.370, 337.120, 347.182, and 348.995\,GHz, with a bandwidth 1.875\,GHz (frequency domain mode), to cover the CO(11-10) line (1267.014486\,GHz at $z\!=\!0$, or 348.176556\,GHz at $z\!=\!2.639$). The channel width was 0.976563\,MHz and the total channel number was 1920 for each window.

The calibration and imaging of the data were carried out using the Common
Astronomy Software Application package (CASA; \citealt{mcmullin2007}) version 5.1. In order to improve phase coherence, we carried out phase-only self-calibration for our previous Cycle 2 \citep{inoue2017} and new Cycle 4 data, separately.
In each iteration process, we selected continuum channels and the four
spectral windows were combined to have a sufficient signal-to-noise ratio
(S/N $\geq 3.0$) for each gain solution.

After applying obtained calibration solutions to the whole data (continuum
and lines), we carried out imaging with a natural weighting and further
flagged some bad antennas by visual inspection of visibilities.
After phase-only self-calibration, our Cycle 2 data were significantly improved.
The 'dynamic range,' defined as peak intensity per root-mean-square (rms) error,
increased from $\sim 120$ to $\sim 250$.
Applying phase-only self-calibration to our Cycle 4 data turned out to
be difficult due to insufficient dynamic range in the original data. 
However, by reducing the condition for rejecting solutions for a group of three
antennas to S/N $>2.0$, we obtained a dynamic range of $140$, which was
improved by $\sim 20\%$ compared to the original value.

After phase-only self-calibration, we combined the continuum
data with weights inversely proportional to the variance of errors.
In order to assess possible systematic differences in amplitude of
visibilities, we compared both continuum data with a common $uv$-range
between 140 and 1400\,m.
It turned out that the difference is below $\sim 10\%$, which is
comparable to the typical error value in ALMA observations.
We also subtracted the continuum emission from the line channels for both
the Cycles 2 and 4 data in the $uv$ plane and combined them with weights inversely proportional to the variance of errors.
The channel maps of the lensed CO(11-10) emission were generated with a width of 146\,MHz (125.6\,$\textrm{km}\,\textrm{s}^{-1}$) per channel.

\section{Results}

\subsection{Lensed images and line spectrum}
\begin{figure*}
\epsscale{1.2}
\hspace{-0.3cm}
\plotone{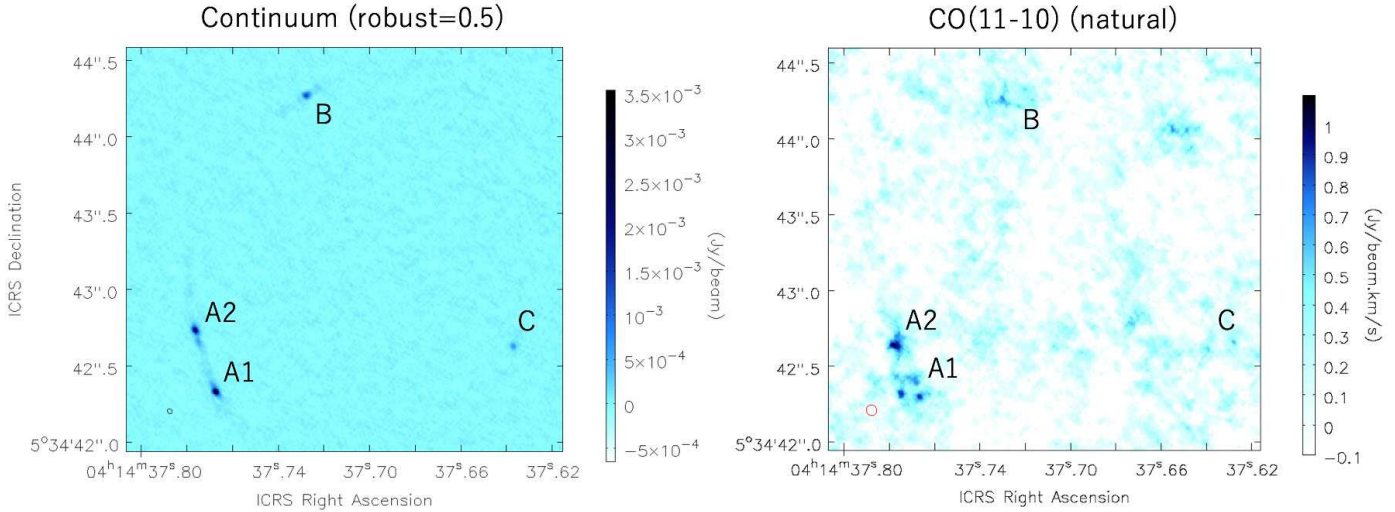}
\caption{ALMA 0.88\,mm (Band 7, 340\,GHz) continuum and CO(11-10) images of MG\,J0414+0534.
The Briggs weighting ($robust=0.5$) continuum image (left) was obtained from the Cycle 4 data, and the natural weighting integrated intensity image of the CO(11-10) (right) was obtained from the combined Cycles 2 and 4 data.
%Imaging has been carried out with a Briggs weighting of $robust=0.5$ for
%the continuum and a natural weighting for the CO(11-10) integrated intensity.
The beam sizes and the position angles measured East of North, which are
shown at the bottom left corner of each panel are
$(0\farcs 036\times 0\farcs 029, 40\fdg 5)$ (left) and
$(0\farcs 072\times 0\farcs 068, 18\fdg 2)$ (right).
The rms noises and the total fluxes of lensed images are
$(32\,\mu \tr{Jy}/\tr{beam}, 16\pm 2\,\tr{mJy})$ (left) and
$(0.19\,\tr{mJy}/\tr{beam}/\tr{channel}, 6.3\pm 0.6\,\tr{Jy}\cdot \tr{km}/\tr{s})$ (right). For making the CO(11-10) image, only signals with $\gtrsim 1\,\sigma$ in each channel map were used.}
\label{f1}
\end{figure*}

Fig.\,\ref{f1} shows the continuum image obtained from the Cycle 4 data and the integrated intensity of the CO(11-10) line emission obtained from the combined Cycles 2 and 4 data.
The continuum and CO images were obtained with a Briggs ($robust=0.5$) and a natural weightings, respectively.
The achieved angular resolutions are $0\farcs03-0\farcs04$ and $\sim 0\farcs 07$, respectively. 
They show bright quadruply lensed spots A1, A2, B, and C. The continuum map shows some arcs on A1, A2, and B. The CO map shows that A1 consists of several spots, suggesting a more complex source structure than that of the continuum. The integrated flux and luminosity of the CO(11-10) line are $6.3\pm 0.6\,\tr{Jy}\cdot \tr{km}/\tr{s}$ and $(1.1 \pm 0.1)\mu^{-1} \times 10^9\,\tr{L}_\odot$, where 
$\mu$ is the total magnification.\footnote{The magnification factor for the mid-infrared emission is expected to be $\mu\sim 40$ \citep{takahashi-inoue2014}.}
\begin{figure}
\epsscale{1.25}
\plotone{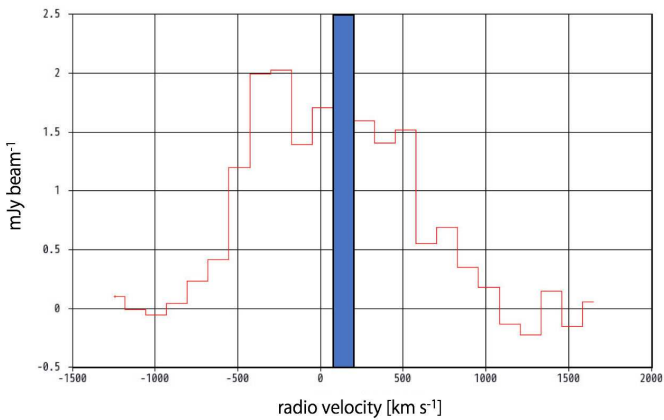}
\caption{The spectrum profile of the continuum subtracted CO(11-10) line emission for A1, A2, and B over the full bandwidth (red). The blue rectangular box at $\sim 140\,$$\textrm{km}\,\textrm{s}^{-1}$ corresponds to missing channels in the Cycle 2 data. The velocity offsets are relative to a redshift of $2.639$. To obtain the spectrum, we smoothed the cube data to have the same angular resolution as in the CO(3-2) data \citep{barvainis1998}.}
\label{f2}
\end{figure}

Fig.\,\ref{f2} shows the line spectrum of CO(11-10) obtained from the combined Cycles 2 and 4 data. We smoothed the cube data using a gaussian $uv$ tapering that yields beam sizes and a position angle of $\sim (2^{\prime\prime}\times 1^{\prime\prime}, 16^\circ)$ at each velocity bin, and only used the areas of the components A1+A2 and B with their integrated intensity $>3 \sigma$ in the lensed plane \citep[the CO(11-10) spectrum is comparable to the CO(3-2) spectrum in][see also Sect.~\ref{sect-dis}]{barvainis1998}.
It shows a broad line width of full width at zero intensity (FWZI) $=1900\,$$\textrm{km}\,\textrm{s}^{-1}$ and FWHM $=1170\pm 180\,$$\textrm{km}\,\textrm{s}^{-1}$ obtained from a Gaussian fit. The dip at $\sim 140\,$$\textrm{km}\,\textrm{s}^{-1}$ is due to missing channels in our Cycle 2 data taken in time domain mode.

\subsection{Lens modeling and de-lensing}
To model the observed images, we first made an unperturbed 'background'
model without using our ALMA data.
It consists of a singular isothermal ellipsoid for a lensing galaxy G 
plus a cored isothermal sphere for a possible companion galaxy, object X \citep[see][]{falco1997}, and an external shear.
For fitting relative positions of the lensed images, we used the measured
\textit{Hubble Space Telescope}(HST) Wide Field and
Planetary Camera/Wide Field and Planetary Camera 2 (WFPC/WFPC2) positions and centroids of lensing galaxies, G and X in
the CASTLES data archive\footnote{http://www.cfa.harvard.edu/castles/}.
For fitting flux ratios of lensed images, we used the mid-infrared (MIR) flux
ratios of A2/A1 and B/A1 \citep{minezaki2009, macleod2013}. The Very Long Baseline 
Interferometry (VLBI) positions of the lensed possible core components at $5\,$GHz \citep{trotter2000} were fit to the corresponding HST positions. The residual errors in the relative positions are $<0\farcs 003$, which are less than the errors in the HST relative positions.

In order to improve our model, we took into account the lensing effects from possible substructures (subhalos and line-of-sight halos/voids) using our ALMA Cycles 2 and 4 combined continuum image (not shown here). We obtained a best-fit model by minimizing the differences between the de-lensed images, while the astrometric shifts and magnification perturbations were constrained to fit the optical/near-infrared (NIR) and MIR data \citep[see][for detail]{inoue2016,inoue2020}. After fitting, the positions of possible core components \textit{p} and \textit{q}, jet components \textit{r} and \textit{s}
at $5\,$GHz \citep[see][for the definition]{trotter2000} and the optical/IR data were fit to the observed data.

Using the best-fit lensing model, we reconstructed de-lensed images of the continuum and CO(11-10) line from a linear combination of three de-lensed images of A1, A2 and B. Image C was not taken into account as the angular resolution of de-lensed image C was too low. From the de-lensed images, we found that the total magnification factors for the continuum and CO(11-10) are  $37\pm4$ and  $28\pm 3$, respectively.

\subsection{Continuum}
\label{sect-cont}
\begin{figure*}
\epsscale{1.21}
\hspace{-0.7cm}
\plotone{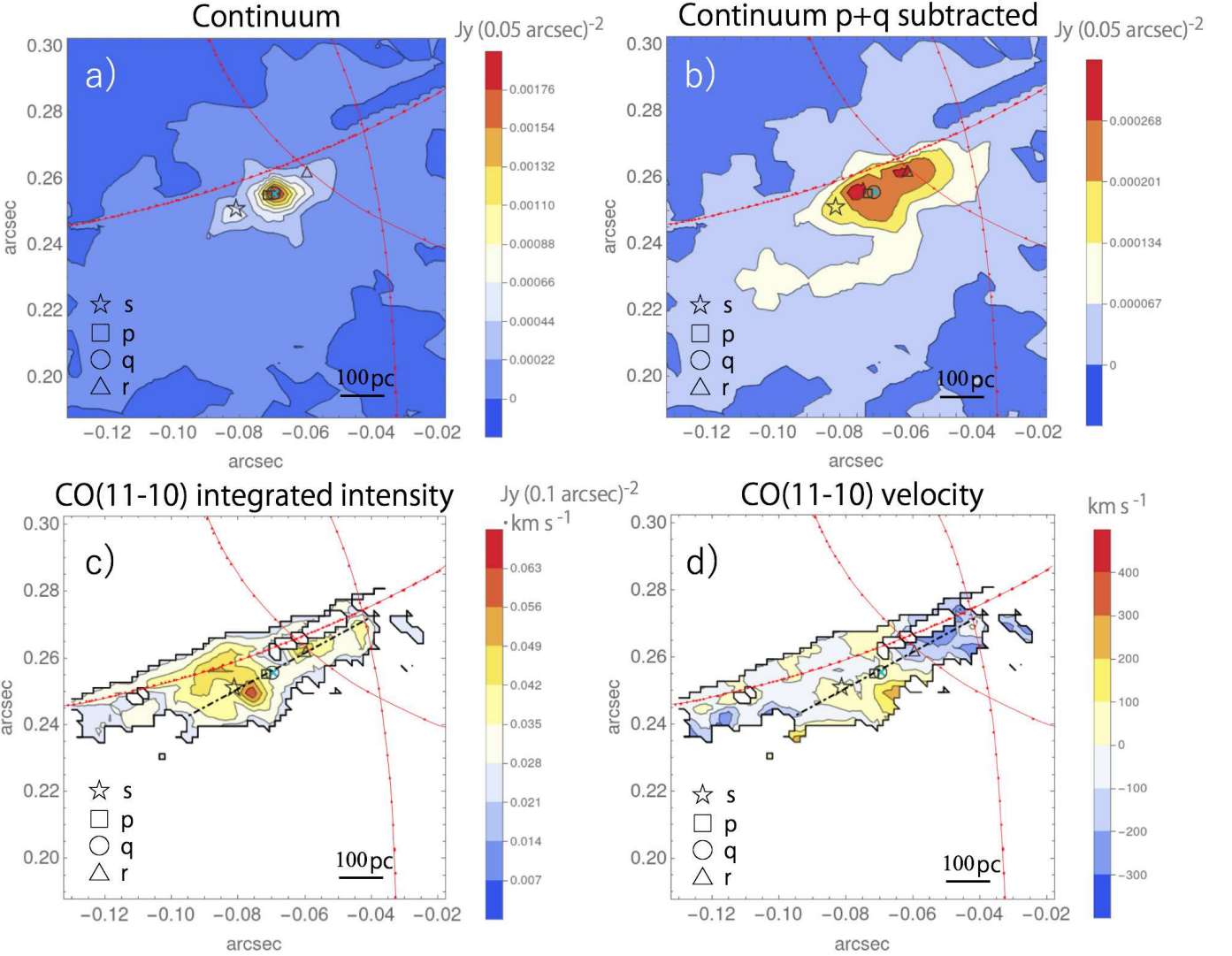}
\caption{
(a) De-lensed continuum image of the Cycle 4 data before subtraction of $\textit{p}+\textit{q}$,
(b) de-lensed continuum image of the Cycle 4 data with $\textit{p}+\textit{q}$ subtracted,
(c) de-lensed CO(11-10) integrated intensity over the full band width, and
(d) de-lensed CO(11-10) intensity weighted velocity maps.
In all the maps, $0\farcs01$ corresponds to a physical scale of 82\,pc. The coordinates are in the FK5 frame at J2000.0 centered at the centroid of a lensing galaxy G. The spatial resolution of the continuum image is $\sim 50$\,pc and that of the CO(11-10) image is $\sim 100$\,pc. The contours start from $5\,\sigma $ for (a) and (b), $3\,\sigma $ for (c). The spacings are  $10\,\sigma $ for (a), $3\,\sigma $ for (b), $1\,\sigma $ for (c) and $100\, \textrm{km}\, \textrm{s}^{-1}$  for (d). For both of the CO(11-10) maps, pixels below the $3\,\sigma$ in the integrated intensity map are masked out, and the limit is expressed as thick contours. The center velocity was shifted by $-75\,$$\textrm{km}\,\textrm{s}^{-1}$ from $z_\ti{S}=2.639$ to make the quasar core \textit{q} be at rest in the intensity weighted velocity map. The green cross in each panel corresponds to the best-fit position of the MIR emission from the quasar dust core. Red dots and fitted curves correspond to the caustics. The dashed-dotted line in the CO(11-10) maps shows a section for the position-velocity (PV) diagram in Fig.\,\ref{f4}.}
\label{f3}
\end{figure*}

Fig.\,\ref{f3}(a) shows the de-lensed continuum image of the Cycle 4 data ($\textit{robust}=0.5$)\footnote{We also made a de-lensed image using our Cycles 2 and 4 combined data. However, the angular resolution of the de-lensed image was slightly worse than the image in Fig.\,\ref{f3}.}. It shows a very
bright compact emission at the position of the core component \textit{q}. 
Moreover, the best-fit source position of the MIR emission (denoted by a 
green cross in Fig.\,\ref{f3}) seems to coincide with that of \textit{q}. Though not conclusive, our result suggests that \textit{q} corresponds to quasar circumnuclear dust emission or synchrotron radiation from the quasar core, whereas another possible core component \textit{p} corresponds to emission from the quasar jet near the core.

In order to further probe the structure near the core, we attempted to reduce the side effect due to the  'side lobes' of \textit{p} and \textit{q} in the source plane as they are extremely bright. To do so, we subtracted off the best-fit Gaussians at the lensed \textit{p} and \textit{q} images.  The Gaussian point spread function (PSF) flux ratios were chosen to be consistent with the VLBI flux ratios of \textit{p} and \textit{q} at 5\,GHz \citep{trotter2000, inoue2020}. The absolute amplitudes of the lensed \textit{p} and \textit{q} at A2 were chosen as free parameters as they have a relatively large flux and separation. The absolute amplitudes were determined appropriately through visual inspection of the smoothness of the image in which the fitted Gaussian PSFs are subtracted off. 
The achieved angular resolution near intensity peak in the source plane was found to be $\sim 0\farcs 007$\footnote{The resolution can be evaluated from the beam size of the Cycle 4 continuum image and the magnification factors in the vicinity of intensity peaks}, which corresponds to $\sim 50$\,pc at $z_\ti{S}=2.639$.

Fig.\,\ref{f3}(b) shows the de-lensed continuum image of Cycle 4 data ($\textit{robust}=0.5$) in which $\textit{p}+\textit{q}$ are subtracted off. In the image, one can clearly see a bimodal structure along jet components. The positions of local peak intensity in the bimodal structure are slightly shifted toward \textit{q} from the best-fit positions of the other jet components \textit{s} and \textit{r}. The origin and the emission mechanism of the bimodal structure is considered to be either synchrotron
emission from the quasar jets or dust emission, or both. The size of the de-lensed continuum emitting region is $0.5-1$\,kpc. We can also see a filamentary structure that looks like a spiral-arm or a stream in the southeast of \textit{q}, which implies clumpy structure in the dust. The length of the 'arm' (or 'stream') is $\sim 400\,$pc, which is approximately a half of dusty substructures observed in luminous submillimeter galaxies \citep{hodge2019}.

\subsection{CO(11-10)}
\label{sect-co}
Fig.\,\ref{f3}(c) and (d) show the de-lensed CO(11-10) integrated intensity and intensity weighted velocity field maps. In order to increase the S/N, we only picked up pixels with S/N $>2$ within the velocity range of $-700<V<800\,\textrm{km}/\textrm{s}$ and combined them. For the velocity definition, we assumed that the core \textit{q} has a zero velocity: the center velocity was shifted by $-75\,$$\textrm{km}\,\textrm{s}^{-1}$ from $z_\ti{S}=2.639$, meaning that the rest frame is at the redshift of the CO emitting core, $z_\tr{CO}=2.6381$. In the de-lensed integrated intensity image (Fig.~\ref{f3}c), no strong core component can be seen, which is contrary to the continuum image that shows a strongest peak in there.
Instead, we can see a bimodal structure along jet components in a similar way as is the continuum image with PSFs subtracted. The positions of the local peak intensity appear to roughly coincide with the brightest regions in the PSFs subtracted continuum image; they reside in the vicinity of jet components \textit{s} and \textit{r}, though slightly shifted toward \textit{q}. In the velocity field map (Fig.~\ref{f3}d), one can see a velocity gradient of $-2\,\textrm{km}/\textrm{s} \textrm{pc}^{-1}$ along a section that connects \textit{q} and \textit{r}. However, it turned out that no such velocity gradient can be seen along a section that connects \textit{q} and \textit{s}. Such an asymmetric feature cannot be explained by a simple circular rotating disk or a spheroid.

\begin{figure}
\epsscale{1.25}
\plotone{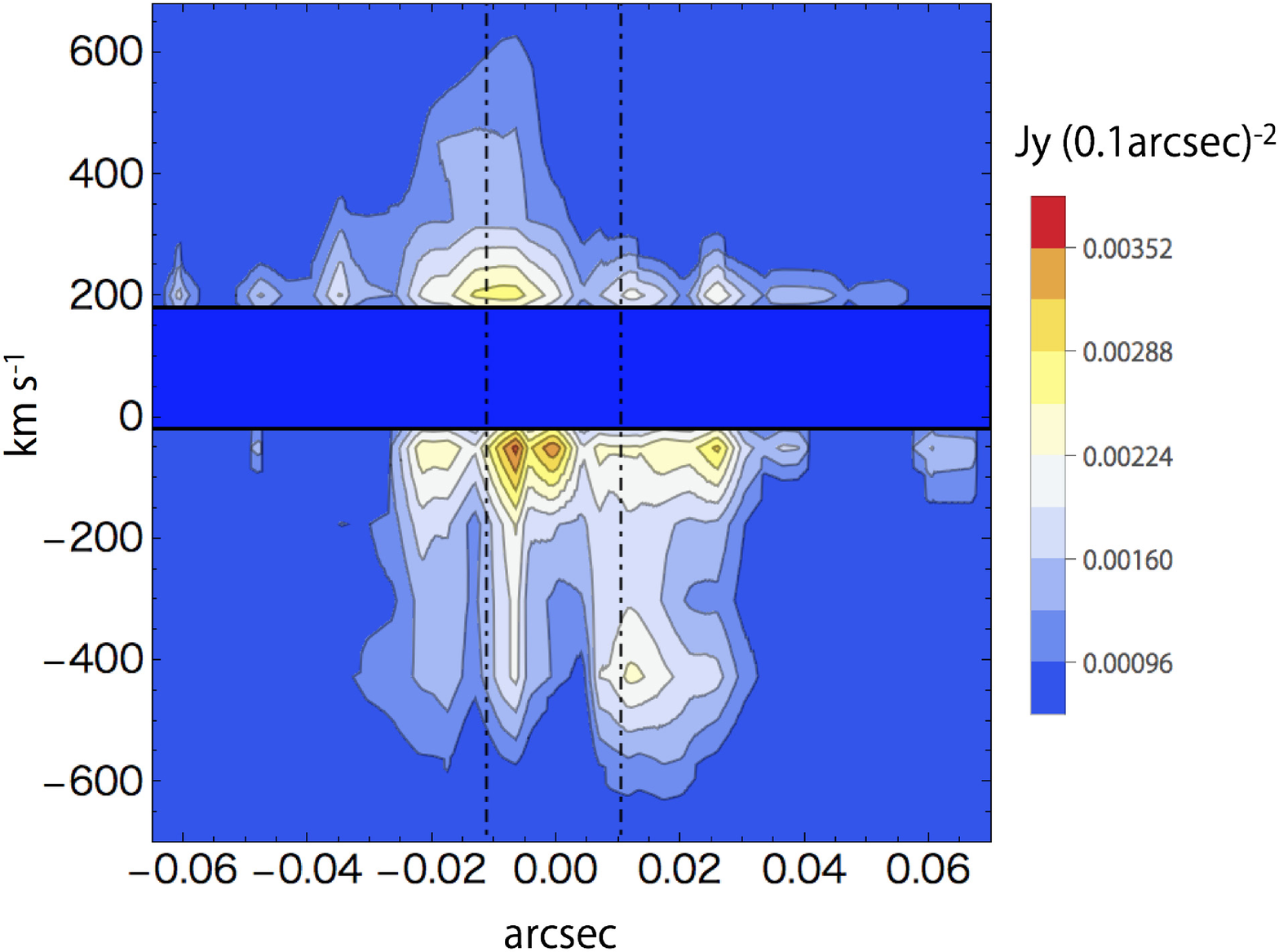}
\caption{
Position-velocity (PV) diagram of the de-lensed CO(11-10) data
on a section that best fits the positions of all the components \textit{p}, \textit{q},
\textit{r}, and \textit{s} (see Fig.\,\ref{f3}(d)). The data around $\sim 140\,$$\textrm{km}\,\textrm{s}^{-1}$ are omitted due to misisng channels in our Cycle 2 observations (see also Fig. 2). The dot-dashed vertical lines 
show the position offsets of jet components \textit{s} (left line) and \textit{r} (right line) with respect to \textit{q} measured in the section. Pixels below the $3\,\sigma$ in intensity are masked out.
The contour spacing is $1\,\sigma$. The zero-velocity point corresponds to the rest frame of the quasar core \textit{q} and the zero-position point corresponds to the position that is nearest to \textit{q} in the section.}
\label{f4}
\end{figure}

In order to inspect possible interaction between jet and gas, we made a position-velocity (PV) diagram (Fig.\,\ref{f4}) on a section that best fits the positions of the three jet components \textit{s}, \textit{p} and \textit{r} and the core component \textit{q} using $\chi^2$ minimization (Fig.\,\ref{f3}d). Data at $\sim 140$\,$\textrm{km}\,\textrm{s}^{-1}$ were omitted due to missing channels in our Cycle 2 'time domain mode' observations. One can see that the brightest spot in the vicinity of the eastern jet component \textit{s} has velocities up to $\pm 600$\,$\textrm{km}\,\textrm{s}^{-1}$. In addition, the CO gas in the vicinity of the western jet component \textit{r} is moving at velocities from $-300$\,$\textrm{km}\,\textrm{s}^{-1}$ to 600\,$\textrm{km}\,\textrm{s}^{-1}$. Both the \textit{s} and \textit{r} components are located at a distance of $\sim 80$\,pc from the quasar core component \textit{q}.
Even some components are moving at velocities from $-400$\,$\textrm{km}\,\textrm{s}^{-1}$ to 400\,$\textrm{km}\,\textrm{s}^{-1}$ at the outer side of \textit{s} and from $-200$\,$\textrm{km}\,\textrm{s}^{-1}$ to 600\,$\textrm{km}\,\textrm{s}^{-1}$ at the outer side of \textit{r}, both with respect to \textit{q}. 

\section{Discussion}
\label{sect-dis}
We detected a broad (FWZI $\sim1900$ $\textrm{km}\,\textrm{s}^{-1}$ and FWHM $\sim1170\pm 180\,$$\textrm{km}\,\textrm{s}^{-1}$), high-J (J = 11-10) CO emission line from the lensed quasar MG\,J0414+0534. This line is much broader than the lower transition CO(3-2) line with FWHM $= 580\,$$\textrm{km}\,\textrm{s}^{-1}$ \citep{barvainis1998}. Since the S/N of the CO(3-2) data is much worse than that of our CO(11-10) data, weak high-velocity components may have not been detected in the CO(3-2) data. Alternatively, the high-velocity components are dominated by highly excited CO and affected by the differential magnification effect \citep{serjeant2012}. Assuming that both the lines except for the high-velocity components are emitting from the same region, restricting the channels of CO(11-10) to have a line width of $ 6\times 10^2\,\textrm{km}\,\textrm{s}^{-1}$, which is equivalent to the FWHM of the CO(3-2) data, the CO(11-10)/(3-2) brightness temperature ratio is found to be 0.14. Since the CO(1-0) line has not been detected, the lower limit of the CO(3-2)/(1-0) line luminosity ratio is estimated as $2.56$ \citep{sharon2016}, and that of the CO(3-2)/(1-0) brightness temperature ratio is estimated as $0.475$. Using RADEX \citep{vandertak2007}, we found that a kinetic temperature of $\sim 2 \times 10^{2}$ K and a $\textrm{H}_2$ number density of $\sim5\times10^{4}$ cm$^{-3}$ are consistent with the brightness temperature ratios of CO. These facts indicate that the molecular gas in MG\,J0414+0534 is in high-temperature and high-density conditions.

Such conditions are likely due to the interaction between the jets and molecular gas. First, our ALMA data show a broad CO(11-10) line emission along the radio jet components. The positions of the bright CO(11-10) components coincide well with those of the bright radio jet components, and those CO(11-10) components have broad line widths (Sect.\,\ref{sect-co}), which are very similar to what has been observed in IC 5063 \citep{morganti2015}. Secondly, the 22 GHz water maser lines are detected from this lensed quasar, with their velocities (the central velocities of two strongest line components are at $-278$ $\textrm{km}\,\textrm{s}^{-1}$ and $+470$ $\textrm{km}\,\textrm{s}^{-1}$) well within the broad CO(11-10) line width, and it is suggested that the water maser lines come from the interaction zone of the radio jet and the molecular gas \citep{impellizzeri2008,castangia2011}.

The high-temperature and high-density conditions are also consistent with the characteristics of the continuum radio emission. The spectral energy distribution (SED) of the continuum radio emission of MG\,J0414+0534 has a turnover around $0.4-0.5$ GHz, and therefore MG\,J0414+0534 is classified as a giga-peaked spectrum (GPS) source \citep{katz1997,edwards2004}. The major cause of the SED turn-over in GPS sources is considered as either synchrotron self-absorption or free-free absorption \citep[see][for a review]{odea1998}. Together with the small size ($\lesssim 1\,$kpc) of their radio emitting regions, it implies that radio jets in GPS sources are still interacting with the ISM in the host galaxies \citep{odea1997}. The compactness of the continuum emitting region of $500-1000$\,pc (Sect.\,\ref{sect-cont}) and the molecular gas $-$ jet interaction we discussed above in MG\,J0414+0534 are well consistent with the characteristics of GPS sources.

It is suggested that GPS sources are in the young stage of the evolution of radio galaxies on the grounds of their small sizes \citep{odea1998} as seen in MG\,J0414+0534. On the other hand, GPS sources might be so frustrated that the expansions of jets are regulated by the dense ISM, and the condition could make the sizes of the GPS sources relatively small with respect to their old ages \citep{odea1997}. Nonetheless, the ages of GPS sources are supposed to be as young as $\lesssim 1 \times 10^{5}$ years even under the frustrated conditions.
Recent numerical simulations of jet-ISM interaction show that even though dense ISM is surrounding radio sources, jets can expand to $\sim 1000$ pc within the order of $10^{4}$ years \citep{wagner2011,wagner2012}. On the other hand, intermittent radio jet activity can be a counterargument against young age of GPS sources. However, there is no observational evidence that supports intermittent radio jet activity in MG\,J0414+0534. It does not have any extended radio emission, and also most of GPS sources do not have any extended radio emission \citep{odea1998}.

Based on the evidence above, we conclude that MG\,J0414+0534 is possibly in the early stage of quasar radio activity. We expect that the radio jets recently emitted from the AGN is currently drilling the ISM inside the host galaxy. The interaction between the radio jets and the ISM in the quasar host galaxy causes the highly excited broad CO(11-10) line and the water maser lines.

Recent observations toward AGNs and quasars detected highly excited CO lines \citep[e.g.,][]{hailey-dunsheath2012}. The jet-ISM interaction may be responsible for heating the CO gas. Our observations would provide an important insight on a possible excitation mechanism for high-J CO lines.

\acknowledgments
KTI would like to thank Eiji Akiyama, Misato Fukagawa and Fumi Egusa for their support on data reduction,
and acknowledges support from NAOJ ALMA Scientific Grant Number 2018-07A
and JSPS KAKENHI Grant Number 17H02868.
SM is supported by the Ministry of Science and Technology (MoST)  of Taiwan,
MoST 103-2112-M-001-032-MY3, 106-2112-M-001-011, and 107-2119-M-001-020.  
KN is supported by JSPS KAKENHI Grant Number 19K03937.
This Letter makes use of the following ALMA data:
ADS/JAO.ALMA$\#$2013.1.01110.S., 2016.1.00281.S. ALMA is a partnership of ESO (representing its member states), NSF (USA) and NINS (Japan), together with NRC (Canada) 
and NSC and ASIAA (Taiwan), and KASI (Republic of Korea) in
cooperation with the Republic of Chile. The Joint ALMA Observatory is
operated by ESO, AUI/NRAO, and NAOJ.
%\appendix
%
%\section{Appendix information}

\bibliographystyle{aasjournal}
\bibliography{jet-dust2019}
%\begin{thebibliography}{}
\expandafter\ifx\csname natexlab\endcsname\relax\def\natexlab#1{#1}\fi
\providecommand{\url}[1]{\href{#1}{#1}}
%\end{thebibliography}{}

\end{document}